\documentclass[a4paper,12pt]{amsart}
\usepackage{amssymb,amsmath,amscd}
\usepackage{url}
\usepackage[all,cmtip]{xy}
\usepackage{graphicx}
\usepackage{tikz}
\usepackage{pgfplots}
\usepackage{enumerate}
\usetikzlibrary{matrix,arrows,decorations.pathmorphing}

\usepackage[mathlines,pagewise]{lineno}

\usepackage{graphicx}

\newtheorem{theorem}{Theorem}[section]
\newtheorem{lemma}[theorem]{Lemma}
\newtheorem{proposition}[theorem]{Proposition}
\newtheorem{corollary}[theorem]{Corollary}

\newtheorem{remark}[theorem]{Remark}
\newtheorem{example}[theorem]{Example}
\newtheorem*{theorem*}{Theorem}
\newtheorem*{lemma*}{Lemma}
\newtheorem*{proposition*}{Proposition}
\newtheorem*{corollary*}{Corollary}

\newtheorem*{remark*}{Remark}

\newcommand{\be}{\begin{eqnarray}}
\newcommand{\ee}{\end{eqnarray}}

\begin{document}

\title{A Class of Automatic Sequences}

\author{Michel Rigo \and Robert Underwood}

\address{Institut de Math\'{e}matiques \\
Universit\'{e} de Li\`{e}ge \\
Belgium}
\email{m.rigo@uliege.be}
\address{Department of Mathematics and Computer Science \\
Auburn University at Montgomery  \\
Alabama, USA}
\email{runderwo@aum.edu}

\date{\today}

\maketitle

\begin{abstract} Let $k\ge 2$.  We prove that the characteristic sequence of a regular language over a $k$-letter alphabet is $k$-automatic.  More generally, if $t\ge 2$ and $t,k$ are multiplicatively dependent, we show that the characteristic sequence of a regular language over a $t$-letter alphabet is $k$-automatic. 
\end{abstract}

{\it keywords:}\ finite automaton; regular language; automatic sequence


\section{Introduction}   
Automatic sequences were first developed by Alan Cobham as uniform tag sequences arising from uniform tag systems in the general sense \cite[Theorem 3]{Co72}, \cite[Theorem 6.3.2]{AS03}, \cite[Section 2.3]{Ri14}.   For $k\ge 2$, a $k$-automatic sequence is defined as the image of a fixed point of a morphism on the internal symbol set of a uniform tag system of 
modulus $k$ (such morphisms are also called $k$-uniform prolongable morphisms).   In this paper, we are concerned with the following problem:\ Let $t\ge 2$, let $\Sigma$ be a $t$-letter alphabet, let $L\subseteq \Sigma^*$ be a regular language, and let $\{s_{L,r}\}_{r\ge 0}$ be the characteristic sequence of $L$ with respect to the geneological ordering of $\Sigma^*$.   
Under what conditions on $t,k$ is $\{s_{L,r}\}_{r\ge 0}$ $k$-automatic?

In general, the characteristic sequence of a regular language over a $t$-letter alphabet is {\em not} $k$-automatic.  Indeed, take $t=2$ and $k=3$, and let $\Sigma=\{a,b\}$ and $L=a^*$.   Then the characteristic sequence is 
\[\{s_{L,r}\}_{r\ge 0} = 1\,1\,0\,1\,0\,0\,0\,1\,0\,0\,0\,0\,0\,0\,0\,1\dots\]
which is $2$-automatic.  To see this, let $\langle \{q_1,q_2,q_3\},q_1,w,h,\{0,1\}\rangle$ be the uniform
tag system of modulus $2$ with $w(q_1)=q_1q_2$, $w(q_2)=q_3q_2$, $w(q_3)=q_3q_3$, and 
$h(q_1)=h(q_2)=1$, $h(q_3)=0$.  Then $\{s_{L,r}\}=h(\lim_{n\rightarrow \infty} w^n(q_1))$, so that $\{s_{L,r}\}$ is $2$-automatic. Now,
$\{s_{L,r}\}$ is also the characteristic sequence of the set of non-negative integers $\{2^n-1\}_{n\ge 0}$, and so, 
$\{2^n-1\}_{n\ge 0}$ is $2$-recognizable \cite[p. 187]{Co69}.  However, $\{2^n-1\}_{n\ge 0}$ is not $3$-recognizable, for if so, then by \cite[Theorem, p. 186]{Co69}, $\{2^n-1\}$ is ultimately periodic, which is not the case.
Thus $\{s_{L,r}\}$ is not $3$-automatic.
 
Returning to the general case, it remains to determine which values of $t,k$ yield $\{s_{L,r}\}$ $k$-automatic.   In our main result (Theorem \ref{main-result}) we show that if $t=k$, then $\{s_{L,r}\}$ is $k$-automatic.   As a corollary (Corollary \ref{cor}), we also show that if $t,k$ are multiplicatively dependent, then $\{s_{L,r}\}$ is $k$-automatic.

\section{Tag Systems and Automatic Sequences}

In this section we introduce uniform tag systems and show how they determine automatic sequences.  We retain the original
terminology of Cobham.

A {\em tag system} is a $3$-tuple ${\mathcal T}=\langle B,b_1,w\rangle$ where
$B$ is a finite set (the {\em internal symbol set}), $b_1$ is an element of $B$ (the {\em initial symbol}), and $w$ is a function 
from $B$ to $B^*$ (the {\em production function}) which satisfies $w(b_1)=b_1y_2y_3\dots y_n$, that is, the first symbol of
$w(b_1)$ is $b_1$.  The production function can be extended to words in $B^*$ as follows: for $y_1y_2\dots y_n\in B^*$,
\[w(y_1y_2\dots y_n)=w(y_1)w(y_2)\dots w(y_n).\]
From this, we can define non-negative powers of $w$:\ for $b\in B$, $n\in {\mathbb N}$, 
\[w^0(b)=b,\quad w^{n+1}(b)=w(w^n(b)).\]
A {\em tag system in the general sense} is a $5$-tuple ${\mathcal T}'=\langle B,b_1,w,h,A\rangle$ 
where ${\mathcal T}=\langle B,b_1,w\rangle$ is a tag system, and where $h$ is a function from $B$ to a finite set $A$ (the {\em external symbol set}).  A tag system is {\em uniform of modulus $k$} if there exists an integer $k\ge 2$ for which $w(b)$ is a word of length $k$ for each $b\in B$.

\begin{proposition} Let ${\mathcal T}=\langle B,b_1,w,h,A\rangle$ be a uniform tag system of modulus $k$.  Let 
\[b=\lim_{n\rightarrow \infty} w^n(b_1).\]
Then $b$ is the unique infinite word over $B$ that begins with $b_1$ and is a fixed point of $w$, that is, $w(b)=b$. 
\end{proposition}

\begin{proof}   The conditions on $b$ follow since ${\mathcal T}$ is uniform of modulus $m\ge 2$.  See \cite[p. 167, second paragraph]{Co72}. 
\end{proof}

The infinite word 
\[b=\lim_{n\rightarrow  \infty} w^n(b_1)=b_1y_2y_3y_4y_5y_6\dots\]
is the {\em internal sequence} of ${\mathcal T}$, denoted as $\mathrm {intseq}({\mathcal T})$, and its image
\[h(b)=h(\lim_{n\rightarrow \infty} w^n(b_1))=h(b_1)h(y_2)h(y_3)h(y_4)h(y_5)h(y_6)\dots\]
is the {\em external sequence} of ${\mathcal T}$ denoted as $\mathrm{extseq}({\mathcal T})$. 

\begin{example}[Thue-Morse Tag System]  The $5$-tuple ${\mathcal T}=\langle \{0,1\},0,w,I,\{0,1\}\rangle$ with $w(0)=0\,1$, 
$w(1)=1\,0$, and $I:\{0,1\}\rightarrow \{0,1\}$ the identity function, is a uniform tag system of 
modulus $2$.  One has
\[\mathrm {intseq}({\mathcal T})=\mathrm {extseq}({\mathcal T})= 0\,1\,1\,0\,1\,0\,0\,1\,1\,0\,0\,1\,0\,1\,1\,0\,1\,0\,0\,1\,0\,1\,1\,0\,0\,\dots\]
\end{example}
The Thue-Morse external sequence can be described as follows: for $n\ge 0$, the $n$th term of $\mathrm {extseq}({\mathcal T})$ is $1$ if the number of 1's in the $2$-ary representation of $n$ is odd, and is $0$ otherwise.   

\begin{example}[Regular Paperfolding Tag System]  The $5$-tuple 
\[{\mathcal T}=\langle \{q_1,q_2,q_3,q_4\},q_1,w,h,\{0,1\}\rangle\] with 
$w(q_1)=q_1\,q_2$, $w(q_2)=q_1\,q_3$, $w(q_3)=q_4\,q_3$, $w(q_4)=q_4\,q_2$, and $h:\{q_1,q_2,q_3,q_4\}\rightarrow \{0,1\}$ defined 
as $h(q_1)=h(q_2)=1$, $h(q_3)=h(q_4)=0$, is a uniform tag system of 
modulus $2$.  One has
\[\mathrm {intseq}({\mathcal T})= q_1\,q_2\,q_1\,q_3\,q_1\,q_2\,q_4\,q_3\,q_1\,q_2\,q_1\,q_3\,q_4\,q_2\,q_4\,q_3\dots\]
and 
\[\mathrm {extseq}({\mathcal T})=1\,1\,1\,0\,1\,1\,0\,0\,1\,1\,1\,0\,0\,1\,0\,0\,1\,1\,1\,0\,1\,1\,0\,0\,\dots\]
\end{example}

The name ``Regular Paperfolding" is appropriate since $\mathrm{extseq}({\mathcal T})$ can be obtained by taking an ordinary piece of paper, making a sequence of \lq\lq up-folds\rq\rq\ towards the left edge, then unfolding.  The resulting sequence of \lq\lq valleys\rq\rq\ ($=1$) and \lq\lq ridges\rq\rq\ ($=0$) gives the sequence, see \cite[\S 1]{CV12}.  (Note: our sequence above has an initial 
term $=1$.) 

\begin{example}[Even Numbers Tag System]  The $5$-tuple ${\mathcal T}=\langle \{0,1\},1,w,I,\{0,1\}\rangle$ with 
$w(0)=0\,1\,0$, $w(1)=1\,0\,1$, and $I(0)=0$, $I(1)=1$, is a uniform tag system of 
modulus $3$.  One has
\[\mathrm {intseq}({\mathcal T})=\mathrm {extseq}({\mathcal T}) =  1\,0\,1\,0\,1\,0\,1\,0\,1\dots\]
\end{example}

Let $k\ge 2$.  A sequence is {\em $k$-automatic} if it is the external sequence $\mathrm{extseq}({\mathcal T})$ of a uniform tag system ${\mathcal T}$ of modulus $k$. 

\begin{example}[Thue-Morse Sequence]  Let ${\mathcal T}=\langle \{0,1\},0,w,I,\{0,1\}\rangle$ be the Thue-Morse tag system of modulus $2$.  Then its external sequence

\[\mathrm {extseq}({\mathcal T})= 0\,1\,1\,0\,1\,0\,0\,1\,1\,0\,0\,1\,0\,1\,1\,0\,1\,0\,0\,1\,0\,1\,1\,0\,0\,\dots\]
is the {\em Thue-Morse $2$-automatic sequence}.
\end{example}

\begin{example}[Regular Paperfolding Sequence]  Let ${\mathcal T}=\langle \{q_0,q_1,q_2,q_3\},q_0,w,h,\{0,1\}\rangle$ be the Regular Paperfolding tag system of modulus $2$.   Then the external sequence
\[\mathrm {extseq}({\mathcal T})=1\,1\,1\,0\,1\,1\,0\,0\,1\,1\,1\,0\,0\,1\,0\,0\,1\,1\,1\,0\,1\,1\,0\,0\,\dots\]
is the {\em Regular paperfolding $2$-automatic sequence}.
\end{example}

\begin{example}[Even Numbers Sequence]  Let ${\mathcal T}=\langle \{0,1\},1,w,I,\{0,1\}\rangle$ be the Even Numbers tag system of modulus $3$.  Then the external sequence
\[\mathrm {extseq}({\mathcal T})=1\,0\,1\,0\,1\,0\,1\,0\,1\dots\]
is the {\em Even numbers $3$-automatic sequence}.
\end{example}

More generally, every ultimately periodic sequence is $k$-automatic for $k\ge 2$ \cite[Theorem 5.4.2]{AS03}.

\section{Cobham's Main Result}

We review the main result of Cobham's 1972 paper \cite[Theorem 3]{Co72}, which shows that 
uniform tag systems and finite automata are essentially equivalent.   

For $k\ge 2$, let $D_k=\{0,1,2,3,\dots,k-1\}$ denote the set of $k$-ary digits.  A {\em finite automaton over $D_k$} is a $4$-tuple ${\mathcal A}=\langle Q,q_1,\delta,{\mathcal F}\rangle$ consisting of a finite set of {\em states} $Q=\{q_1,q_2,\dots,q_n\}$, an {\em initial state} $q_1\in Q$, a {\em transition function} 
$\delta: Q\times D_k\rightarrow Q$, and a partition ${\mathcal F}=\{F_0,F_1,\dots, F_m\}$ of $Q$.  We always assume that $\delta(q_1,0)=q_1$.  We extend $\delta$ by composition to the 
function $\hat \delta: Q\times D_k^*\rightarrow Q$.  

A finite automaton ${\mathcal A}=\langle Q,q_1,\delta,{\mathcal F}\rangle$ over $D_k$ determines two sequences:\ The {\em state sequence} of ${\mathcal A}$ is the sequence
\[\mathrm{state}({\mathcal A}) = \{y_i\} = \{\hat \delta(q_1,\bar i)\},\]
where $\bar i$ is the $k$-ary representation of $i\in {\mathbb N}$.   The {\em sorting  sequence} of ${\mathcal A}$ is the sequence
\[\mathrm{sort}({\mathcal A}) = \{z_i\} = \{F_i\}\quad \hbox{if}\ \hat \delta(q_1,\bar i)\in F_i.\]

\begin{example}  The $4$-tuple ${\mathcal A}=\langle \{q_1,q_2\}, q_1,\delta,\{\{q_1\},\{q_2\}\}\rangle$ is a finite automaton over $D_2=\{0,1\}$ defined by the state diagram given in Fig. 1.
\begin{figure}[t]
\centerline{\hspace{-3.5cm}
$
\qbezier(75,20)(65,40)(80,50)
\qbezier(90,50)(105,40)(95,20)
\qbezier(80,50)(85,52.5)(90,50)
\put(74,23){\vector(1,-3){2}}
\
\put(80,60){$0$}
\put(15,0){\circle{30}}
\put(85,0){\circle{30}}
\
\qbezier(5,20)(-5,40)(10,50)
\qbezier(20,50)(35,40)(25,20)
\qbezier(10,50)(15,52.5)(20,50)
\put(4,23){\vector(1,-3){2}}
\put(47,10){$1$}
\put(35,5){\vector(1,0){30}}
\put(65,-5){\vector(-1,0){30}}
\put(47,-20){$1$}
\put(-24,0){\vector(1,0){20}}
\put(9,-2.5){$q_1$}
\put(79,-2.5){$q_2$}
\put(10,60){$0$}
$}
\vspace{.5cm}
\begin{center}
\small Fig. 1.\ \ State diagram of Example 3.1.
\end{center}
\end{figure}
\normalsize
The state sequence is
\[\mathrm{state}({\mathcal A}) = q_1\,q_2\,q_2\,q_1\,q_2\,q_1\dots\]
and the sorting sequence is 
\[\mathrm{sort}({\mathcal A}) =  \{q_1\}\,\{q_2\}\,\{q_2\}\,\{q_1\}\,\{q_2\}\,\{q_1\}\dots\]
\end{example}

Here is Cobham's main result \cite[Theorem 3]{Co72}.

\begin{theorem}[Cobham]\label{cobham} Let ${\mathcal A}=\langle Q,q_1,\delta,{\mathcal F}\rangle$ be a finite automaton over $D_k$.
Then there exists a uniform tag system ${\mathcal T}=\langle Q,q_1,w,h,{\mathcal F}\rangle$ of modulus $k$ for which
\[\mathrm{intseq}({\mathcal T}) = \mathrm{state}({\mathcal A})\quad \hbox{and}\quad \mathrm{extseq}({\mathcal T}) = \mathrm{sort}({\mathcal A}).\]
Conversely, let  ${\mathcal T}=\langle B,b_1,w,h,A\rangle$ be a uniform tag system of modulus $k$.  Then there exists
a finite automaton ${\mathcal A}=\langle B,b_1,\delta,{\mathcal F}\rangle$ over $D_k$ for which
\[\mathrm{state}({\mathcal A}) =  \mathrm{intseq}({\mathcal T}) \quad \hbox{and}\quad 
\mathrm{sort}({\mathcal A}) = h^{-1}(\mathrm{extseq}({\mathcal T})),\] 
where $h^{-1}(a)=\{b\in B:\ h(b)=a\}$ for $a\in A$.
\end{theorem}

\begin{corollary}  The preimage under $h$ of a $k$-automatic sequence is the sorting sequence of a finite automaton over $D_k$. 
\end{corollary}

\begin{remark}  Regarding Theorem \ref{cobham}, if ${\mathcal A}=\langle Q,q_1,\delta, {\mathcal F}\rangle$ is a finite automaton over $D_k$, then the production function of the corresponding uniform tag system 
${\mathcal T}=\langle Q,q_1,w,h,{\mathcal F}\rangle $ is defined as
\[w(q)=\delta(q,0)\delta(q,1)\delta(q,2)\dots \delta(q,k-1),\]
and the ``output" function $h: Q\rightarrow {\mathcal F}$ is given as $h(q)=F_i$ if $q\in F_i$, for $q\in Q$. 
Conversely, given a uniform tag system ${\mathcal T}=\langle B,b_1,w,h,A\rangle$ of modulus $k$, the transition function
of the corresponding finite automaton is defined as
\[\delta(b,i)=(i+1)st\text{ symbol in } w(b),\]
for $b\in B$, $i\in D_k$.  
\end{remark}

\begin{remark}   Let ${\mathcal T}=\langle B,b_1,w,h,A\rangle$ be a uniform tag system of modulus $k$ with automatic sequence $\mathrm{extseq}({\mathcal T})$. By Theorem \ref{cobham}, there exists
a finite automaton ${\mathcal A}=\langle B,b_1,\delta,{\mathcal F}\rangle$ over $D_k$ so that
$h(\mathrm{sort}({\mathcal A})) = \mathrm{extseq}({\mathcal T})$.  Thus $k$-automatic sequences can be defined in terms of ``finite automata with output".  This is the definition
of automatic sequence that one is likely to see in the current literature \cite{AS03}, \cite{CV12}, \cite{Ro15}.
\end{remark}

\section{Our Main Result}

In this section we prove our main result: we show that the characteristic sequence of a regular language over an alphabet of $k$ letters is $k$-automatic for $k\ge 2$.   To this end, we construct a uniform tag system ${\mathcal T}$ of modulus $k$ whose external sequence $\mathrm{extseq}({\mathcal T})$ is the given characteristic sequence.  This will yield the result since by definition the sequence $\mathrm{extseq}({\mathcal T})$ is $k$-automatic.   As a corollary, we show that if $t\ge 2$ and $t,k$ are multiplicatively dependent, then the characteristic sequence of a regular language over a $t$-letter alphabet is $k$-automatic.

We begin with some preliminaries.  Let $\Sigma_k=\{1,2,\dots,k\}$ be a finite alphabet on $k$ letters. The words in $\Sigma_k^*$ are genealogically ordered, thus,
\be\label{gen-ord}
\{x_r\}_{r\ge 0}:=\varepsilon,1,2,\ldots,k,11,12,\ldots,1k,21, \ldots
\ee
Note that the rank of a word $c_l\cdots c_0\in \Sigma_k^*$ in the above enumeration (starting with $x_0$ for the empty word) is exactly its base-$k$ value:
\be\label{rank-prop}
\text{if }x_r=c_l\cdots c_0, \text{then } r=\sum_{i=0}^l c_i\, k^i.
\ee

A {\em finite state machine over $\Sigma_k$} is a $4$-tuple ${\mathcal FSM}=\langle Q,q_1,\delta,F\rangle$ consisting of a finite set of {\em states} $Q$, an {\em initial state} $q_1$, a {\em transition function} 
$\delta: Q\times \Sigma_k\rightarrow Q$, and a set of {\em final (or accepting) states} $F\subseteq Q$.   
As before, we extend $\delta$ by composition to the function $\hat \delta: Q\times \Sigma_k^*\rightarrow Q$.  
The definition of a finite state machine is very similar to the definition of finite automaton given in Section 3 (one difference:\ in place of the partition ${\mathcal F}$ of the set of states, we have a subset $F$ of final states). 

A language $L$ is {\em  accepted} by an ${\mathcal FSM}$ if 
\[L=\{w\in \Sigma_k^*:\ \hat\delta(q_1,w)\in F\}.\]   
A language is {\em regular} if it is accepted by an ${\mathcal FSM}$.

\begin{example}\label{no-22s}  Let $\Sigma_2=\{1,2\}$ and let $L\subseteq \Sigma_2^*$ be the language consisting of all words of finite length that contain no consecutive $2$'s.   Then $L$ is regular since it is accepted by the finite state machine
${\mathcal FSM}=\langle \{q_1,q_2,q_3\},q_1,\delta,\{q_1,q_2\}\rangle$
given in Fig. 2.
\begin{figure}[b]
\centerline{\hspace{-5cm}
$
\put(150,0){\circle{30}}
\put(143,-2.5){$q_3$}
\
\qbezier(140,20)(130,40)(145,50)
\qbezier(155,50)(170,40)(160,20)
\qbezier(145,50)(150,52.5)(155,50)
\put(139,23){\vector(1,-3){2}}
\
\put(140,60){$1,2$}
\put(10,0){\circle{30}}
\put(80,0){\circle{28}}
\put(10,0){\circle{28}}
\put(80,0){\circle{30}}
\
\qbezier(0,20)(-10,40)(5,50)
\qbezier(15,50)(30,40)(20,20)
\qbezier(5,50)(10,52.5)(15,50)
\put(-1,23){\vector(1,-3){2}}
\
\put(45,10){$2$}
\put(45,-17){$1$}
\put(-29,0){\vector(1,0){20}}
\put(5,-2.5){$q_1$}
\put(75,-2.5){$q_2$}
\put(10,60){$1$}
\put(100,0){\vector(1,0){30}}
\put(113,5){$2$}
\
\put(30,5){\vector(1,0){30}}
\put(60,-5){\vector(-1,0){30}}
$
}
\vspace{.5cm}
\begin{center}
\small Fig. 2:\ Finite state diagram for Example 4.1.  Accepting states are $q_1,q_2$. 
\end{center}
\end{figure}
\normalsize
\end{example}

Let $L\subseteq \Sigma_k^*$ be 
a regular language and let $\rho_L: \Sigma_k^*\rightarrow \{0,1\}$ be the characteristic
function of $L$, that is,
\[
\rho_L(w) = \left\{
\begin{array}{ll}
1 & \mbox{if $w\in L$} \\
0 & \mbox{if $w\not \in L$.}
\end{array}
\right. 
\]
The {\em characteristic sequence of $L$ with respect to the genealogical order (\ref{gen-ord})} is the 
sequence $\{s_{L,r}\}$ defined as $s_{L,r}=\rho_L(x_r),\ r\ge 0$.

\begin{example} \label{seq-no-22s} Let $L\subseteq \Sigma_2^*=\{1,2\}^*$ be the regular language consisting of all words that contain no consecutive $2$'s.  With respect to the genealogical order 
\begin{equation*}
\{x_r\}_{r\ge 0}:=\varepsilon,1,2,11,12,21,22,111,112,121,122,211,212,221,222,1111,\dots
\end{equation*}
the characteristic sequence $\{s_{L,r}\}=\{p_L(x_r)\}$, $r\ge 0$, is  
\begin{equation*}\label{char-example}
1\,1\,1\,1\,1\,1\,0\,1\,1\,1\,0\,1\,1\,0\,0\,1\,1\,1\,0\,1\,1\,0\,0\,1\,1\,\dots
\end{equation*}
\end{example}

Let ${\mathcal FSM}=\langle Q,q_1,\delta,F\rangle$ be a finite state machine accepting a (regular) language $L$ over $\Sigma_k$.  
Let $Q=\{q_1,q_2,\dots,q_m\}$, and let
\[q_1\cdot x_r=\hat\delta(q_1,x_r)\]
denote the halting state of the ${\mathcal FSM}$ upon reading $x_r$ from its initial state $q_1$.
For $r\ge 0$, define 
\begin{eqnarray} \label{gen-form}
w(q_1\cdot {x_r}) = (q_1\cdot {x_{rk}})(q_1\cdot {x_{rk+1}})(q_1\cdot {x_{rk+2}})(q_1\cdot {x_{rk+3}})\dots (q_1\cdot {x_{rk+(k-1)}}).
\end{eqnarray}
For $y_1,y_2,\dots, y_t\in \Sigma_k^*$, define
\[w((q_1\cdot y_1)(q_1\cdot y_2)\cdots (q_1\cdot y_t)) = w(q_1\cdot y_1)w(q_1\cdot y_2)\cdots w(q_1\cdot y_t).\]
Then $w$ is a morphism on the set of finite sequences of elements in $\{q_1\cdot x:\ x\in \Sigma_k^*\}$.  
Moreover, the composition $w^n$, $n\ge 0$, is well-defined and  
\be\label{limit}
\lim_{n\rightarrow \infty} w^n(q_1\cdot {x_0})=(q_1\cdot {x_0})(q_1\cdot {x_1})(q_1\cdot {x_2})(q_1\cdot {x_3})(q_1\cdot {x_4})\dots
\ee

\begin{lemma}\label{1-fold} Let $x_ix_j$ denote concatenation in $\Sigma_k^*$.  For $r>0$,
\[w(q_1\cdot {x_r}) = (q_1\cdot x_{r-1}x_k)(q_1\cdot x_rx_1)(q_1\cdot x_rx_2)(q_1\cdot x_rx_3)\dots (q_1\cdot x_rx_{k-1}).\]
\end{lemma}

\begin{proof}  By formula (\ref{gen-form})
\[w(q_1\cdot {x_r}) = (q_1\cdot {x_{rk}})(q_1\cdot {x_{rk+1}})(q_1\cdot {x_{rk+2}})(q_1\cdot {x_{rk+3}})\dots (q_1\cdot {x_{rk+(k-1)}}).\] 
Let $r=c_0+c_1k+c_2k^2+\cdots + c_lk^l$, for $c_i\in \Sigma_k$.   If $c_i\not = 1$ for some $0\le i\le l$, then there exist
unique $c_0',c_1',c_2',\dots,c_l'\in \Sigma_k$ with $r-1 = c_0'+c_1'k+c_2'k^2+\cdots + c_l'k^l$.  Now,
\begin{eqnarray*}
rk  & =  &  (r-1)k+k \\
 & = &  k+ c_0'k+c_1'k^2+c_2'k^3+\cdots + c_l'k^{l+1} \\
\end{eqnarray*}
and so, $x_{rk}=x_{r-1}x_k$.   If $c_i=1$ for all $0\le i\le l$, then
$r-1  = k+k^2+k^3+\cdots + k^l$. Thus
\begin{eqnarray*}
rk  & =  &  (r-1)k+k \\
 & = &  k+ k^2+k^3+\cdots + k^{l+1} \\
\end{eqnarray*}
and so, $x_{rk}=x_{r-1}x_k$.  In either case, $q_1\cdot x_{rk} = q_1\cdot x_{r-1}x_k$.

For $1\le j\le k-1$,
\[rk+j  =  j+ c_0k+c_1k^2+c_2k^3+\cdots + c_lk^{l+1}\]
thus $x_{rk+j} = x_rx_j$, and so, $q_1\cdot x_{rk+j} = q_1\cdot x_rx_j$.    It follows that 
\[w(q_1\cdot x_r) = (q_1\cdot x_{r-1}x_k)(q_1\cdot x_rx_1)(q_1\cdot x_rx_2)(q_1\cdot x_rx_3)\dots (q_1\cdot x_rx_{k-1}).\]
\end{proof}

Define an equivalence relation on the set $\Sigma_k^+=\{x_r:\ r>0\}$ by the rule
\[x_r\sim x_s\text{\ if and only if\ } q_1\cdot x_{r-1}=q_1\cdot x_{s-1}\text{\ and\ }q_1\cdot x_{r}=q_1\cdot x_{s}.\]

\begin{lemma}\label{sim-prop}  If $x_r\sim x_s$, then $w(q_1\cdot x_r)=w(q_1\cdot x_s)$.  
\end{lemma}

\begin{proof}  If $x_r\sim x_s$, then $q_1\cdot x_{r-1}=q_1\cdot x_{s-1}$ and $q_1\cdot x_{r}=q_1\cdot x_{s}$.  By Lemma \ref{1-fold}, 
\[w(q_1\cdot {x_r}) = (q_1\cdot x_{r-1}x_k)(q_1\cdot x_rx_1)(q_1\cdot x_rx_2)(q_1\cdot x_rx_3)\dots (q_1\cdot x_rx_{k-1})\]
and 
\[w(q_1\cdot {x_s}) = (q_1\cdot x_{s-1}x_k)(q_1\cdot x_sx_1)(q_1\cdot x_sx_2)(q_1\cdot x_sx_3)\dots (q_1\cdot x_sx_{k-1}).\]
The states $q_1\cdot x_{r-1}x_k$ and $q_1\cdot x_{s-1}x_k$ are the same since the halting states after reading the words
$x_{r-1}$ and $x_{s-1}$ are the same.  Likewise, the states $q_1\cdot x_{r}x_j$ and $q_1\cdot x_{s}x_j$ are identical
for $1\le j\le k-1$.  It follows that $w(q_1\cdot x_r)=w(q_1\cdot x_s)$. 
\end{proof}
 
\begin{lemma}  The equivalence relation $\sim$ has finite index.  
\end{lemma}

\begin{proof}  Recall that there are $\vert Q\vert<\infty$ states in the ${\mathcal FSM}$ that accepts $L$.  Suppose
that $x_r\not \sim x_s$.  Then either $q_1\cdot x_r\not = q_1\cdot x_s$ or both $q_1\cdot x_r = q_1\cdot x_s$ and 
$q_1\cdot x_{r-1}\not = q_1\cdot x_{s-1}$ hold.  Now, the first case can happen in at most $\vert Q\vert^2-\vert Q\vert$ ways.   The second case can happen in at most $\vert Q\vert(\vert Q\vert^2-\vert Q\vert)=\vert Q\vert^3-\vert Q\vert^2$ ways.  Thus there are at most 
\[\vert Q\vert^2-\vert Q\vert+\vert Q\vert^3-\vert Q\vert^2 = \vert Q\vert^3-\vert Q\vert\]
equivalence classes over $\sim$. 
\end{proof}

Let $[x_r]$ denote the equivalence class containing $x_r$.   Let $[\varepsilon]$ denote the class containing $\varepsilon$, only. Put
$B=\{[x_r]:\ r\ge 0\}$.  Then $\vert B\vert \le \vert Q\vert^3-\vert Q\vert +1$. 

\begin{lemma} \label{v-morphism} Define $v: B\rightarrow B^*$ as 
\[v([\varepsilon]) = [\varepsilon][x_1][x_2]\dots [x_{k-1}],\]
and for $r>0$,
\[v([x_r]) = [x_{kr}][x_{kr+1}][x_{kr+2}]\dots [x_{kr+(k-1)}],\]
and define $v: B^*\rightarrow B^*$ as 
\[v([y_1][y_2]\dots [y_t]) = v([y_1])v([y_2])\dots v([y_t]),\]
 for $[y_1],[y_2],\dots,[y_t]\in B^*$.  Then $v$ is a morphism on $B^*$. 
\end{lemma}

\begin{proof}  We need to show that $v$ is well-defined on equivalence classes.  To this end, suppose that
$x_r\sim x_s$.  By Lemma \ref{sim-prop}, $w(q_1\cdot x_r)=w(q_1\cdot x_s)$, and so, $q_1\cdot x_{rk+i}=q_1\cdot x_{sk+i}$ 
for $0\le i\le k-1$, thus $x_{kr+i}\sim x_{sk+i}$ for
$1\le i\le k-1$.   It remains to show that $x_{rk}\sim x_{sk}$.  But we already have $q_1\cdot x_{rk}=q_1\cdot x_{sk}$. 
Observe that $x_{rk-1}=x_{r-1}x_{k-1}$ since $rk-1 = (r-1)k+k-1$.  Likewise, $x_{sk-1} = x_{s-1}x_{k-1}$.  Thus $q_1\cdot x_{rk-1}=q_1\cdot x_{sk-1}$ since $q_1\cdot x_{r-1}=q_1\cdot x_{s-1}$.  Consequently, $x_{rk}\sim x_{sk}$, and so,
$v([x_r])=v([x_s])$. 
\end{proof}

Here is our main result.

\begin{theorem}  \label{main-result} Let $k\ge 2$.  The characteristic sequence of a regular language over an alphabet of $k$ letters is $k$-automatic.
\end{theorem}

\begin{proof}   Let $L\subseteq \Sigma_k^*$ be a regular language, let ${\mathcal FSM}=\langle Q,q_1,\delta,F\rangle$ be a finite state machine that accepts $L$, and let $\{s_{L,r}\}_{r\ge 0}$ be the characteristic sequence of $L$.   Let $B=\{[x_r]:\ r\ge 0\}$ be the finite set of equivalence classes as above, and let $v: B^*\rightarrow B^*$ be the morphism as in Lemma \ref{v-morphism}.  Since $x_r\sim x_s$ implies $q_1\cdot x_r= q_1\cdot x_s$, there is a well-defined
morphism $f: B\rightarrow \{q_1\cdot x_r:\ r\ge 0\}$ given by $f([x_r])=q_1\cdot x_r$, $r\ge 0$.  Now, 
\[f(\lim_{n\rightarrow \infty}v^n([\varepsilon])) = \lim_{n\rightarrow \infty} w^n(q_1\cdot \varepsilon),\] 
and so,
\[f(\lim_{n\rightarrow \infty}v^n([\varepsilon])) = (q_1\cdot \varepsilon)(q_1\cdot x_1)(q_1\cdot x_2)(q_1\cdot x_3)\dots \]
by formula (\ref{limit}).   Since $x_r\sim x_s$ implies $q_1\cdot x_r= q_1\cdot x_s$, there is a function $h: B\rightarrow \{0,1\}$ 
on equivalence classes defined as
\[
h([x_r]) = \left\{
\begin{array}{ll}
1 & \mbox{if $q_1\cdot x_r\in F$} \\
0 & \mbox{if $q_1\cdot x_r\not \in F$.}
\end{array}
\right. 
\]
Now, $\langle B,[\varepsilon], v,h,\{0,1\}\rangle$ is a uniform tag system of modulus $k$ with
\begin{eqnarray*}
h(\lim_{n\rightarrow \infty} v^n([\varepsilon])) & = &  h([\varepsilon])h([x_1])h([x_2])h([x_3])\dots  \\
& = & \{s_{L,r}\}.
\end{eqnarray*}
\end{proof}

\begin{corollary} \label{cor}  Suppose $t,k\ge 2$ are multiplicatively dependent.   The characteristic sequence of a regular language over an alphabet of $t$ letters is $k$-automatic.
\end{corollary}

\begin{proof}  Suppose $k^m=t^n$ for some $m,n\ge 1$ and let $\{s_{L,r}\}$ be the characteristic sequence of a regular language over a $t$-letter alphabet.  By Theorem \ref{main-result}, $\{s_{L,r}\}$ is $t$-automatic, and so, by \cite[Theorem 6.6.4]{AS03}, $\{s_{L,r}\}$is $t^n$-automatic.  Consequently, $\{s_{L,r}\}$ is $k^m$-automatic,
and again by \cite[Theorem 6.6.4]{AS03}, it is $k$-automatic. 
\end{proof}

\begin{example}\label{tag-no-22s}  Let $\Sigma_2=\{1,2\}$ and let $L\subseteq \Sigma^*$ be the regular language consisting of all words of finite length that contain no consecutive $2$'s (Example \ref{no-22s}).   The characteristic sequence 
$\{s_{L,r}\}$ is  
\begin{equation*}\label{char-example}
1\,1\,1\,1\,1\,1\,0\,1\,1\,1\,0\,1\,1\,0\,0\,1\,1\,1\,0\,1\,1\,0\,0\,1\,1\,\dots
\end{equation*}
In this case, $\vert Q\vert =3$ and there are $9$ equivalence classes under $\sim$,
\[B=\{[\varepsilon],[x_1],[x_2],[x_3],[x_6],[x_7],[x_{13}],[x_{14}],[x_{21}]\}\]
One has 
\[w([\varepsilon]))= [\varepsilon][x_1],\quad  w([x_1])=[x_2][x_3],\quad w([x_2])=[x_2][x_3]\]
\[w([x_3]))= [x_6][x_7],\quad  w([x_6])=[x_2][x_{13}],\quad w([x_7])=[x_{14}][x_7]\]
\[w([x_{13}]))= [x_6][x_{14}],\quad  w([x_{14}])=[x_{14}][x_{14}],\quad w([x_{21}])=[x_6][x_{14}],\]
and $h: B\rightarrow \{0,1\}$ is given as 
\[h([\varepsilon])=h([x_1])=h([x_2])=h([x_3])=h([x_7])=1,\]
\[h([x_6])=h([x_{13}])=h([x_{14}])=h([x_{21}])=0.\]
As one can check, ${\mathcal T}=\langle B,[\varepsilon],w,h,\{0,1\}\rangle$ is a uniform tag system of modulus $2$   
whose internal sequence is
\[\mathrm{intseq}({\mathcal T})=
[\varepsilon]\,[x_1]\,[x_2]\,[x_3]\,[x_2]\,[x_3]\,[x_6]\,[x_7]\,[x_2]\,[x_3]\,[x_6]\,[x_7]\,[x_2]\,[x_{13}]\,[x_{14}]\,[x_7]\dots\]
and whose external sequence is 
\[\mathrm{extseq}({\mathcal T})=\{s_{L,r}\} = 1\,1\,1\,1\,1\,1\,0\,1\,1\,1\,0\,1\,1\,0\,0\,1\dots\]
Thus $\{s_{L,r}\}$ is $2$-automatic.

Cobham's theorem (Theorem \ref{cobham}) says that the uniform tag system 
${\mathcal T}=\langle B,[\varepsilon],w,h,\{0,1\}\rangle$ corresponds to a finite automaton 
${\mathcal A} = \langle B,[\varepsilon],\delta,{\mathcal F}\rangle$ over $D_2=\{0,1\}$
with $h(\mathrm{sort}({\mathcal A})) = \mathrm{extseq}({\mathcal T})$.   Consequently, the $2$-automatic sequence 
$\{s_{L,r}\}=\mathrm{extseq}({\mathcal T})$ can be computed by the finite automaton with output given in Fig. 3.

\begin{figure}[t]
\centerline{\hspace{-7cm}
$
\put(110,-70){\circle{27}}
%\put(110,-50){\circle{22}}
\put(170,-70){\circle{27}}
\put(230,-70){\circle{27}}
%\put(170,-50){\circle{22}}
\put(50,-70){\circle{27}}
%\put(50,-50){\circle{22}}
\put(-40,0){\circle{27}}
%\put(-40,0){\circle{20}}
\
\put(41,-72){$\scriptstyle{[x_7]/1}$}
\put(97,-72){$\scriptstyle{[x_{13}]/0}$}
\put(157,-72){$\scriptstyle{[x_{14}]/0}$}
\put(217,-72){$\scriptstyle{[x_{21}]/0}$}
\ 
\put(20,0){\circle{27}}
%\put(20,0){\circle{20}}
\put(80,0){\circle{27}}
%\put(80,0){\circle{20}}
\put(140,0){\circle{27}}
\put(200,0){\circle{27}}
%\put(140,0){\circle{20}}
\
%\qbezier(-40,65)(-50,85)(-35,95)
%\qbezier(-25,95)(-10,85)(-20,65)
%\qbezier(-35,95)(-30,97.5)(-25,95)
%\put(-41,68){\vector(1,-3){2}}
\ 
%\put(-62,80){$\scriptstyle{0,1}$}
%\put(-37,15){\vector(1,4){5}}
%\put(-45,25){$\scriptstyle{0}$}
\
%\put(-30,50){\circle{27}}
%\put(-34,48){$\scriptstyle{[0]}$}
\
\put(-24,0){\vector(1,0){30}}
\put(36,0){\vector(1,0){30}}
\put(96,0){\vector(1,0){30}}
\put(156,0){\vector(1,0){30}}
\put(-14,5){$\scriptstyle{1}$}
\put(-50,50){$\scriptstyle{0}$}
\put(45,5){$\scriptstyle{0}$}
\put(105,5){$\scriptstyle{1}$}
\put(165,5){$\scriptstyle{0}$}
\
\put(10,-1){$\scriptstyle{[x_1]/1}$}
\put(70,-1){$\scriptstyle{[x_2]/1}$}
\put(130,-1){$\scriptstyle{[x_3]/1}$}
\put(190,-1){$\scriptstyle{[x_6]/0}$}
%\put(117,30){$\scriptstyle{1}$}
\
%\put(148,-15){\vector(1,-2){10}}
\put(120,-30){$\scriptstyle{1}$}
\
\put(130,-14){\vector(-4,-3){65}}
\put(55,-24){$\scriptstyle{1}$}
\
%\put(40,-33){\vector(-1,2){10}}
%\put(25,-27){$\scriptstyle{0}$}
\
%\put(90,-45){\vector(-1,0){20}}
%\put(70,-55){\vector(1,0){20}}
%\put(80,-40){$\scriptstyle{0}$}
%\put(80,-68){$\scriptstyle{1}$}
\
%\put(125,-50){\vector(1,0){30}}
%\put(137,-62){$\scriptstyle{1}$}
\
%\put(215,-50){\vector(-1,0){30}}
%\put(137,-62){$\scriptstyle{1}$}
\ 
\put(160,-100){$\scriptstyle{0,1}$}
\qbezier(160,-85)(150,-105)(160,-115)
\qbezier(180,-85)(190,-105)(180,-115)
\qbezier(160,-115)(170,-122)(180,-115)
\put(159,-88){\vector(1,3){2}}
\ 
\put(-48,-1.5){$\scriptstyle{[\varepsilon]/1}$}
\put(-70,0){\vector(1,0){15}}
\
\qbezier(-50,15)(-60,35)(-45,45)
\qbezier(-35,45)(-20,35)(-30,15)
\qbezier(-45,45)(-40,47.5)(-35,45)
\put(-51,18){\vector(1,-3){2}}
\
\put(65,50){$\scriptstyle{0}$}
\
\qbezier(70,15)(60,35)(75,45)
\qbezier(85,45)(100,35)(90,15)
\qbezier(75,45)(80,47.5)(85,45)
\put(69,18){\vector(1,-3){2}}
\
%\qbezier(130,15)(120,35)(135,45)
%\qbezier(145,45)(160,35)(150,15)
%\qbezier(135,45)(140,47.5)(145,45)
%\put(129,18){\vector(1,-3){2}}
\
\qbezier(40,-85)(30,-105)(40,-115)
\qbezier(60,-85)(70,-105)(60,-115)
\qbezier(40,-115)(50,-122)(60,-115)
\put(39,-88){\vector(1,3){2}}
\put(20,-100){$\scriptstyle{1}$}
\
\qbezier(96,15)(138,50)(180,15)
\put(96,14.5){\vector(-3,-4){2}}
\put(130,40){$\scriptstyle{0}$}
\qbezier(36,-15)(78,-50)(120,-15)
\put(120,-14.5){\vector(3,4){2}}
\
\qbezier(70,-85)(112,-120)(154,-85)
\put(154,-85){\vector(3,4){2}}
\put(115,-100){$\scriptstyle{0}$}
\
\put(190,-14){\vector(-4,-3){65}}
\put(123,-53){\vector(4,3){62}}
\put(152,-50){$\scriptstyle{1}$}
\put(160,-17){$\scriptstyle{0}$}
\
\put(125,-70){\vector(1,0){30}}
\put(215,-70){\vector(-1,0){30}}
\put(130,-78){$\scriptstyle{1}$}
\put(200,-78){$\scriptstyle{1}$}
\
\put(224,-55){\vector(-1,3){13}}
\put(225,-30){$\scriptstyle{0}$}
$
}
\vspace{.5cm}
\begin{center}
\small Fig. 3:\ Finite automaton with output that computes the\break  $2$-automatic sequence $\{s_{L,r}\}$ of Example \ref{tag-no-22s}. 
\end{center}
\end{figure}
\end{example}

\end{document}